\documentclass{article}

\usepackage[pdftex]{hyperref}
\usepackage{subfigure}
\usepackage{graphicx}
\usepackage{amsmath}
\usepackage{natbib}
\usepackage{helvet,soul}
\usepackage[font=small]{caption}
\usepackage{xcolor}
\usepackage[english,spanish]{babel}
\usepackage{authblk}

\begin{document}

\title{Study of the molecular gas towards the N11 region in the Large Magellanic Cloud}

\author[1]{M. Celis Pe\~na}
\author[2]{M. Rubio}
\author[1]{S. Paron}
\author[3]{C. Herrera}
\affil[1]{Instituto de Astronom\'ia y F\'isica del Espacio (IAFE), Argentina}
\affil[2]{Universidad de Chile, Chile}
\affil[3]{Institut de Radioastronomie Millimétrique, France}

\maketitle

\selectlanguage{english} 

\begin{abstract}

We study three subregions in the HII region N11 which is located at the northeast side of the Large Magellanic Cloud (LMC). We used $^{12}$CO and $^{13}$CO J=3--2 data observed with the Atacama Submillimeter Telescope Experiment (ASTE) with an angular and spectral resolution of 22$^{\prime\prime}$ and 0.11 km s$^{-1}$ respectively. From the $^{12}$CO J=3--2 and $^{13}$CO J=3--2 integrated maps we estimated, assuming local thermodynamic equilibrium (LTE), masses in about $10^4$ M$_\odot$ for the molecular clouds associated with each subregion. Additionally, from the mentioned maps we study the $^{12}$CO /$^{13}$CO integrated ratios for each subregion, obtaining values between 8 and 10.

\end{abstract}

\selectlanguage{spanish} 
\begin{abstract}

Estudiamos tres subregiones en la regi\'on HII N11, la cual est\'a ubicada en la parte noreste de la Nube Mayor de Magallanes. Usamos datos del $^{12}$CO y $^{13}$CO J=3--2 obtenidos con el Atacama Submillimeter Telescope Experiment (ASTE) con una resoluci\'on angular y espectral de 22$^{\prime\prime}$ y 0.11 km s$^{-1}$ respectivamente. De los mapas de la emisi\'on integrada del $^{12}$CO y $^{13}$CO estimamos las masas considerando equilibrio termodin\'amico local (LTE), obteniendo $10^4$ M$_\odot$ para las nubes moleculares asociadas a cada subregi\'on. Adicionalmente usando los mapas mencionados se estudiaron los cocientes $^{12}$CO/$^{13}$CO de las l\'ineas integradas para cada subregi\'on, obteni\'endose valores entre 8 y 10.

\end{abstract}

\selectlanguage{english}


\section{Introduction}

The Magalleanic Clouds are excellent laboratories to study star formation on different conditions from the Milky Way. In the Large Magellanic Cloud (LMC), the metalicity is Z $\simeq$ 0.5Z$_\odot$ \citep{KellerWood2006} and the gas-to-dust ratio is a factor of $\sim$ 4 higher than in our galaxy. LMC is located at 50 kpc from us \citep{Persson2004}, and is seen nearly face-on with an inclination angle of $\sim$ 35$^{\circ}$. 

The N11 complex, at the northeast side of the LMC, is one of the most important star forming regions in that galaxy. It has a ring morphology with a central cavity with 170 pc in diameter. This region was previously studied by \citep{Herrera2013}, using  $^{12}$CO J=1-0 and $^{12}$CO J=2-1 lines.

\begin{figure}[!t]
  \centering
  \includegraphics[width=0.9\textwidth]{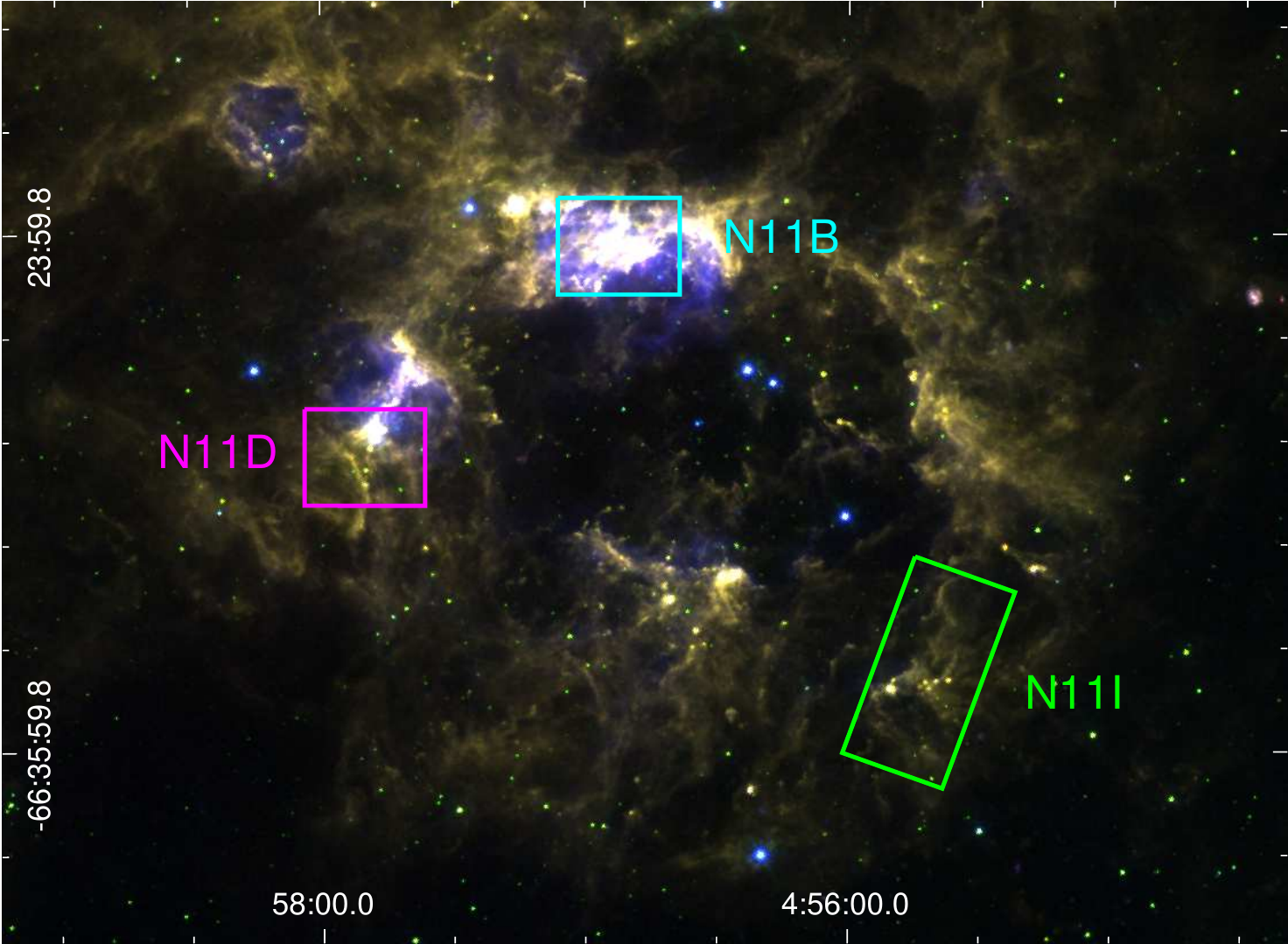}
  \caption{Three-colour image where the 4.5, 5.8 and 8 $\mu$m emission are presented in blue, green and red respectively. The cyan, magenta and green boxes shows the region mapped in the molecular lines.}
  \label{intro}
\end{figure}

\section{Observations}

The observations of the $^{12}$CO $J$=3--2 emission line were performed in October 2014 and the $^{13}$CO $J$=3--2 in August 2015 with the 10 m ASTE telescope. We used the CATS345 GHz band receiver, a two single-band SIS receiver, which is remotely tunable in the LO frequency range of 324-372 GHz. The XF digital spectrometer was set to a bandwidth and spectral resolution of 128 MHz and 125 kHz. The spectral velocity resolution was 0.11 km s$^{-1}$ and the half-power beamwidth (HPBW) was 22$^{\prime\prime}$ at 345 GHz. The system temperature was in the range 150 to 250 K and the main beam efficiency was $\eta_{mb}\sim0.65$. The data were reduced with NOSTAR and the spectra processed using the XSpec software package. The typical rms noise levels are: 0.17 K and 0.11 K for the $^{12}$CO and $^{13}$CO data, respectively.

Figure \ref{intro} shows the N11 region displayed in a three-colour image, in which the 4.5, 5.8, and 8 $\mu$m emission obtained from IRAC-Spitzer are presented in blue, green, and red, respectively. The rectangles represent the regions observed with ASTE: N11B, N11D, and N11I in cyan, magenta, and green boxes, respectively.

\section{Results}

\begin{figure}[!ht]
  \centering
  \includegraphics[width=0.45\textwidth]{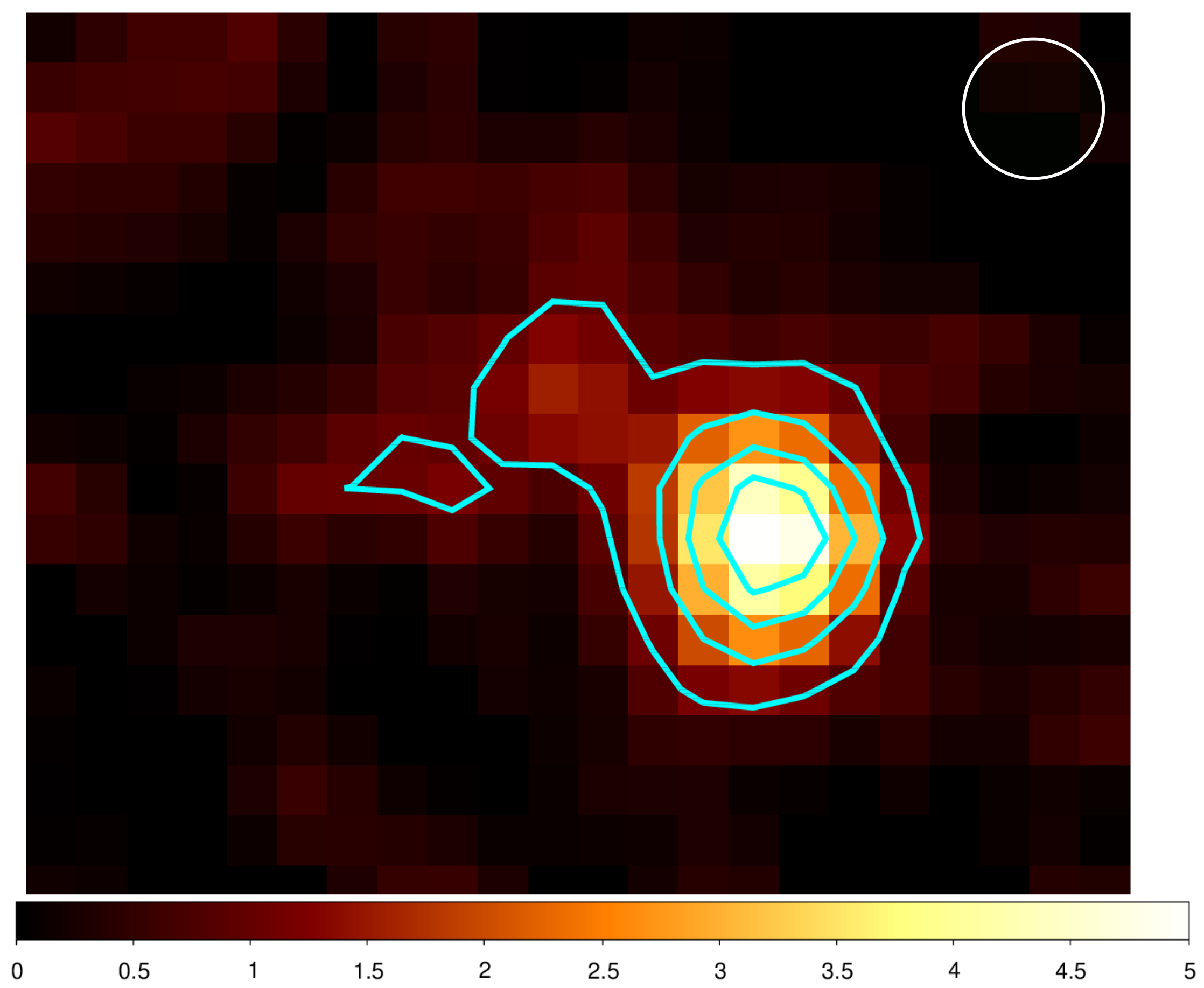}\includegraphics[width=0.48\textwidth]{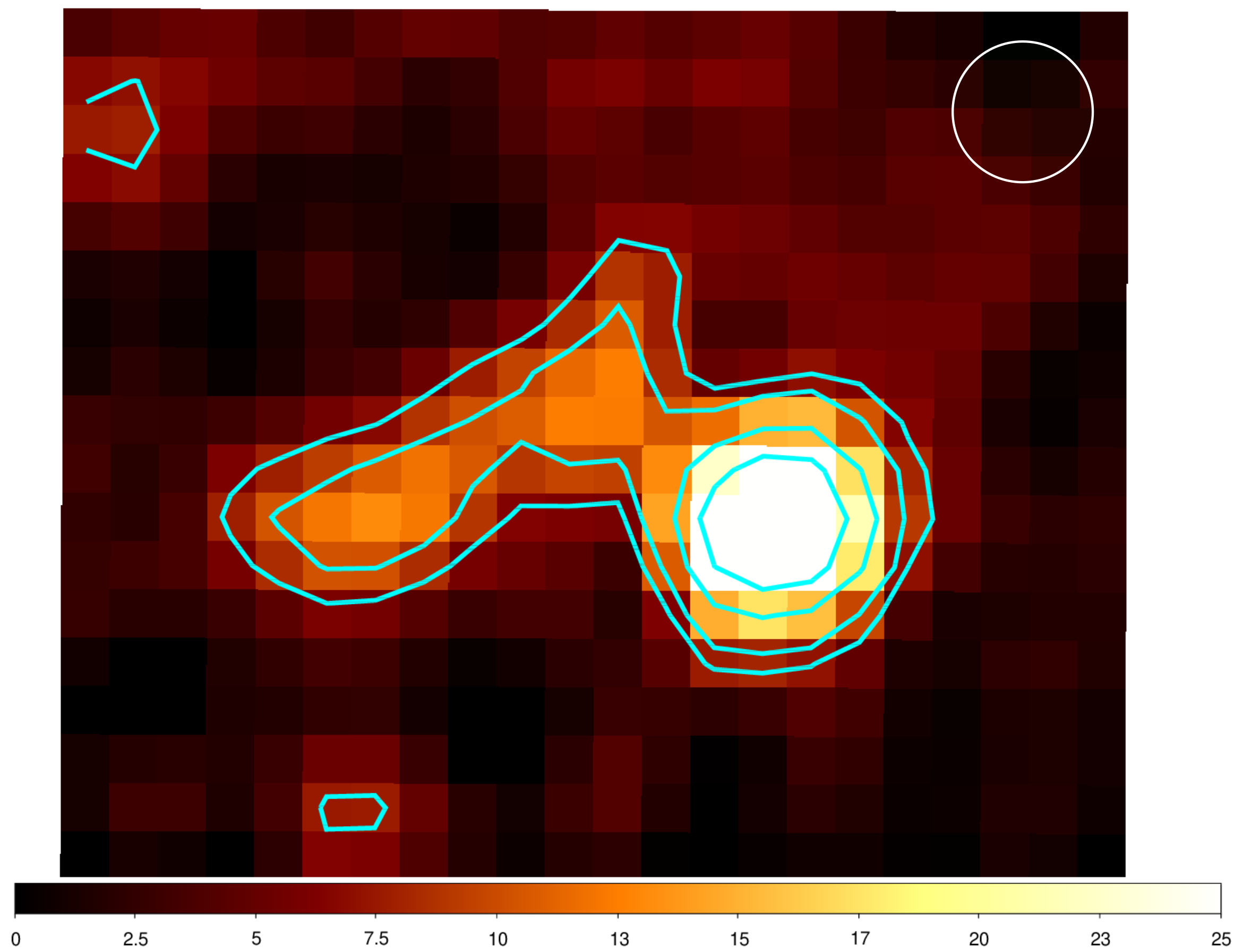}  
  \caption{Region N11B. \textit{Left}: map of $^{13}$CO J=3--2 with contour levels 1, 2, 3 and 4 K km s$^{-1}$.
   \textit{Right}: map of $^{12}$CO J=3--2 with contour levels 7, 10, 17 and 25 K km s$^{-1}$. The beam of the observations is shown at the top right corner.}
  \label{mapasN11B}
\end{figure}

\begin{figure}[!ht]
  \centering
  \includegraphics[width=0.45\textwidth]{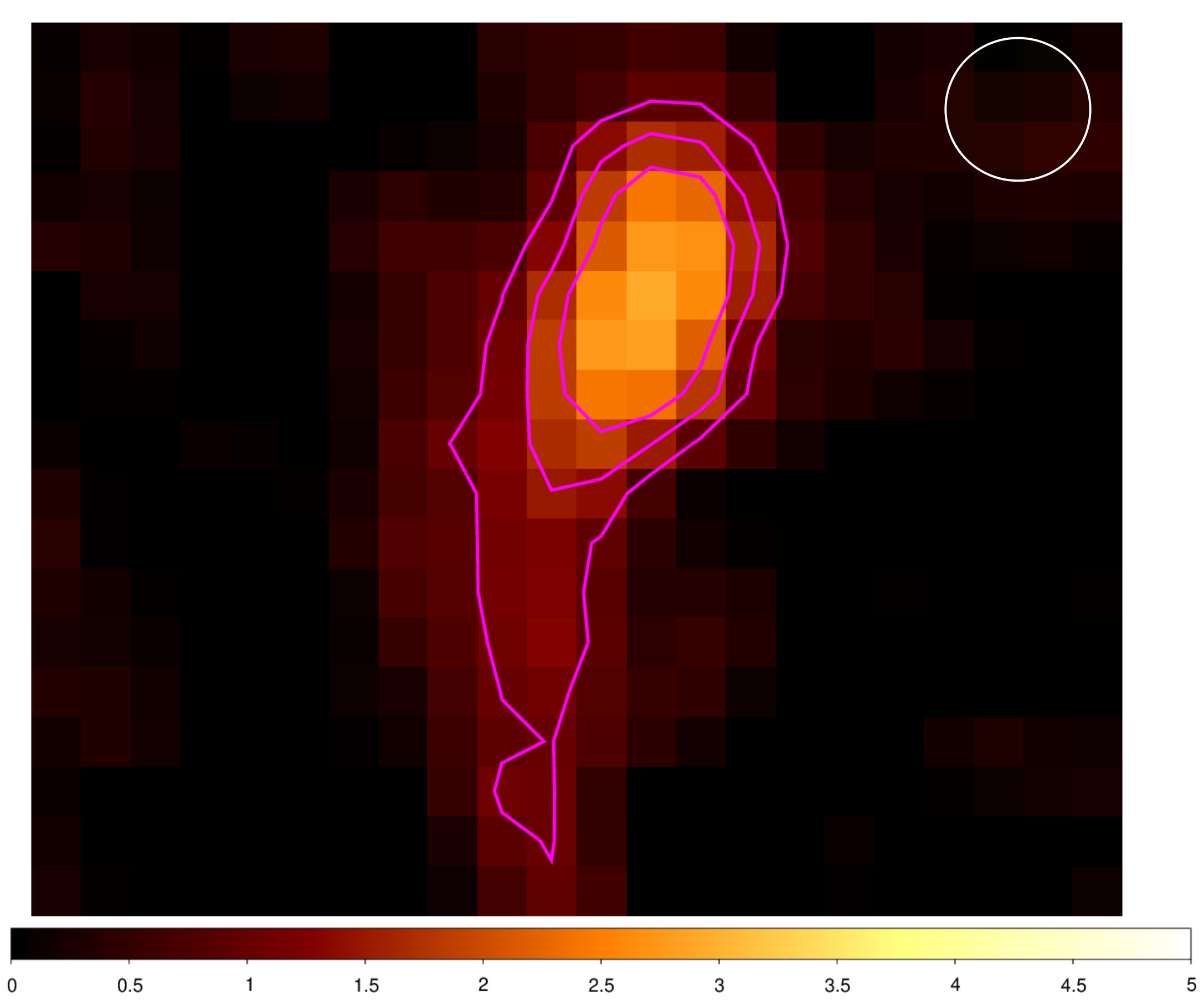}\includegraphics[width=0.45\textwidth]{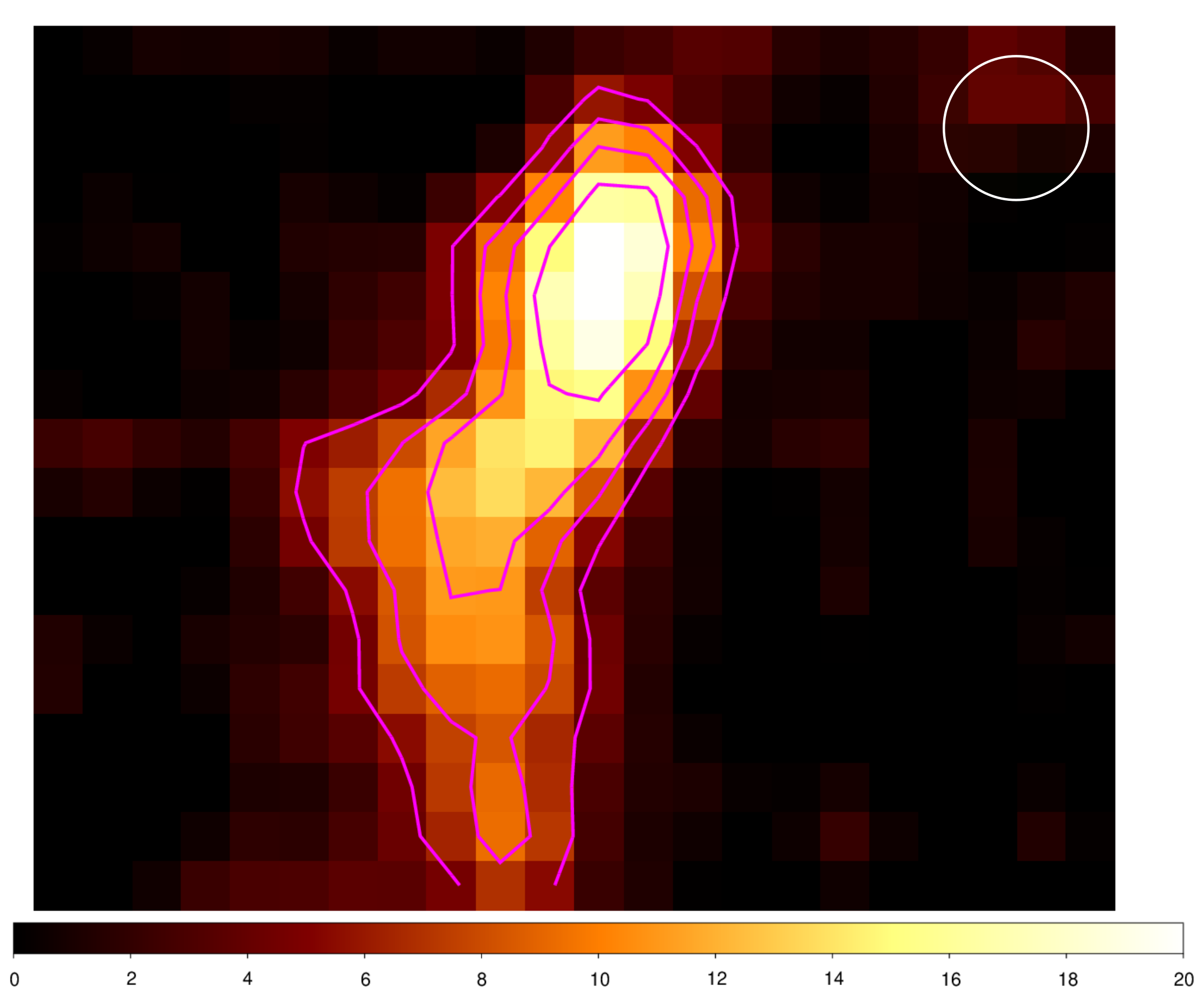}  
  \caption{Region N11D. \textit{Left}: map of $^{13}$CO J=3--2 with contour levels 1, 1.5 and 2 K km s$^{-1}$.
   \textit{Right}: map of $^{12}$CO J=3--2 with contour levels 5, 8, 11 and 15 K km s$^{-1}$. The beam of the observations is shown at the top right corner.}
  \label{mapasN11D}
\end{figure}

\begin{figure}[!ht]
  \centering
  \includegraphics[width=0.45\textwidth]{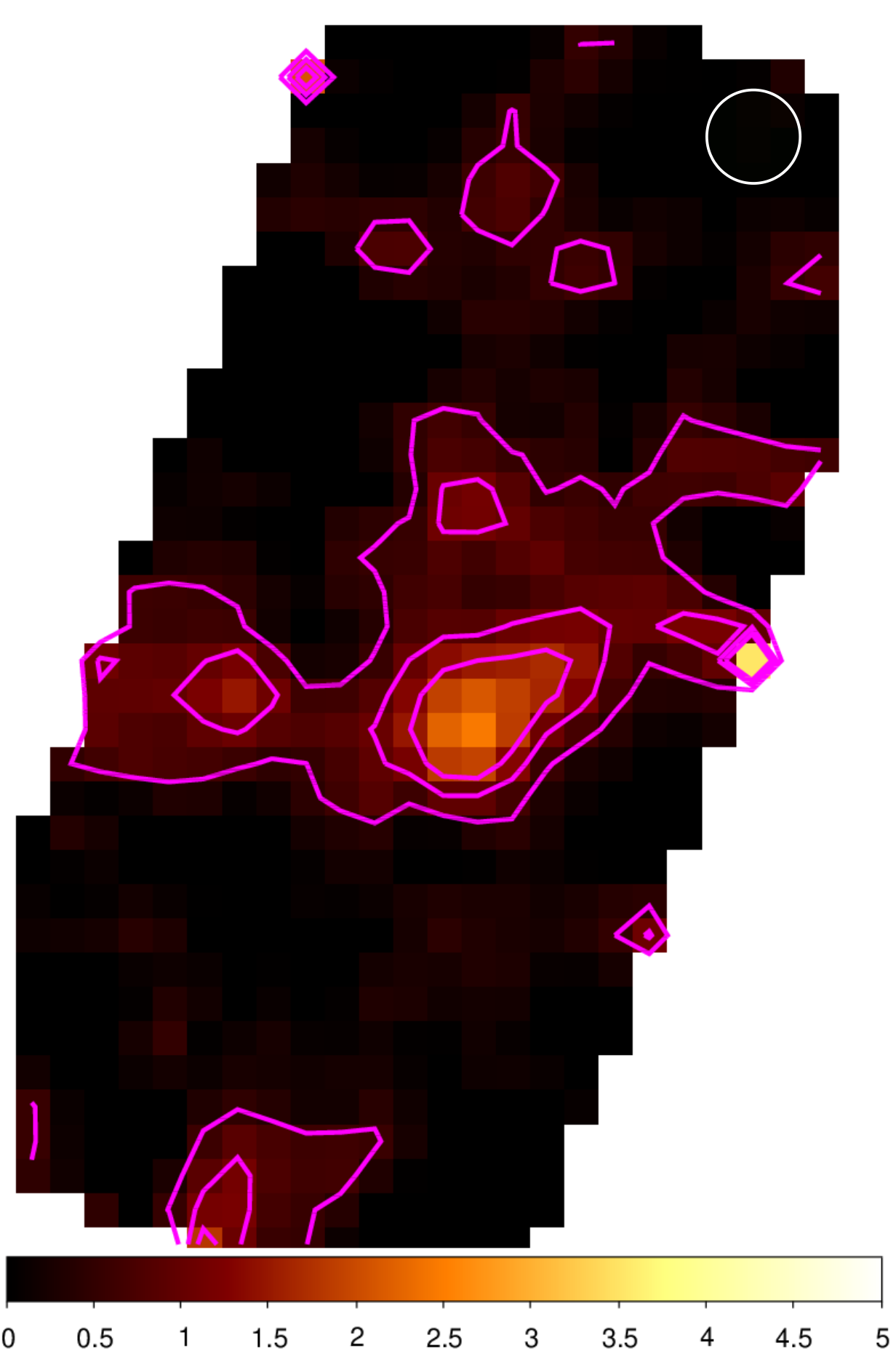}\includegraphics[width=0.45\textwidth]{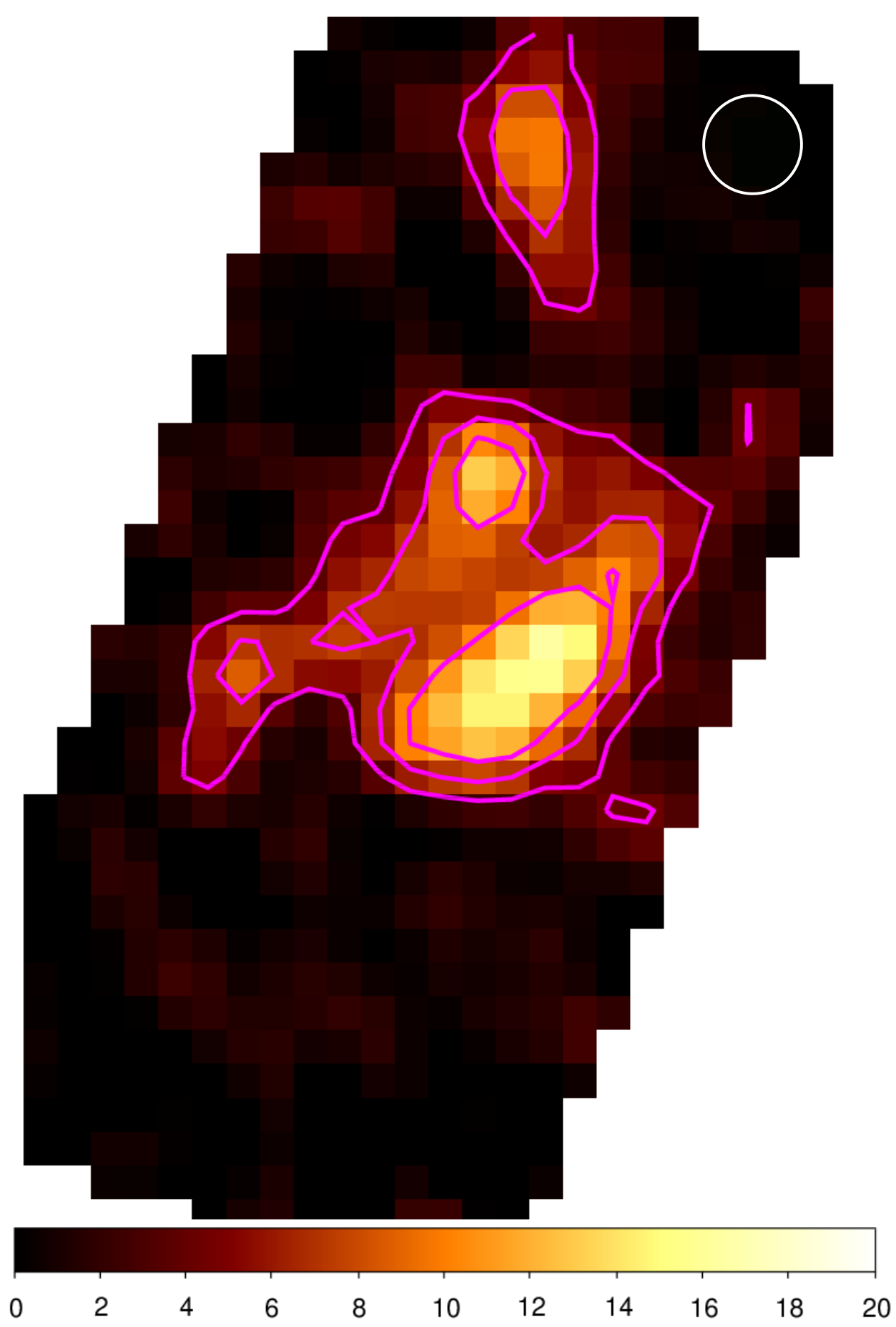}  
  \caption{Region N11I. \textit{Left}: map of $^{13}$CO J=3--2 with contour levels 0.5, 1 and 1.5 K km s$^{-1}$.
   \textit{Right}: map of $^{12}$CO J=3--2 with contour levels 5, 7 and 11 K km s$^{-1}$. The beam of the observations is shown at the top right corner.}
  \label{mapasN11I}
\end{figure}

As shown in Figure \ref{intro} three subregions were studied towards N11. N11B have a size of about 5 pc and it is excited by the OBLH10 star cluster. The observed area was 140$^{\prime\prime}\times$110$^{\prime\prime}$. N11D, which is being excited by the OBLH13 star cluster, has a size of about 9 pc and the observed area was 130$^{\prime\prime}\times$100$^{\prime\prime}$. Finally, it was observed an area of 80$^{\prime\prime}\times$230$^{\prime\prime}$ towards N11I, which has a size of about 10 pc and none star cluster is related to it. Figures \ref{mapasN11B}, \ref{mapasN11D} and \ref{mapasN11I} show the $^{12}$CO and $^{13}$CO J=3--2 integrated between 275 and 295 km s$^{-1}$ for each region, respectively.

As a first result we observe a good morphological and spectral correspondence between the $^{12}$CO J=2--1 emission mapped by \citet{Herrera2013} and the $^{12}$CO and $^{13}$CO J=3--2 presented here. This allow us to perform a detailed multiline analysis of the CO emission towards the region (work in preparation).

\begin{figure}[!ht]
  \centering
  \includegraphics[width=0.7\textwidth]{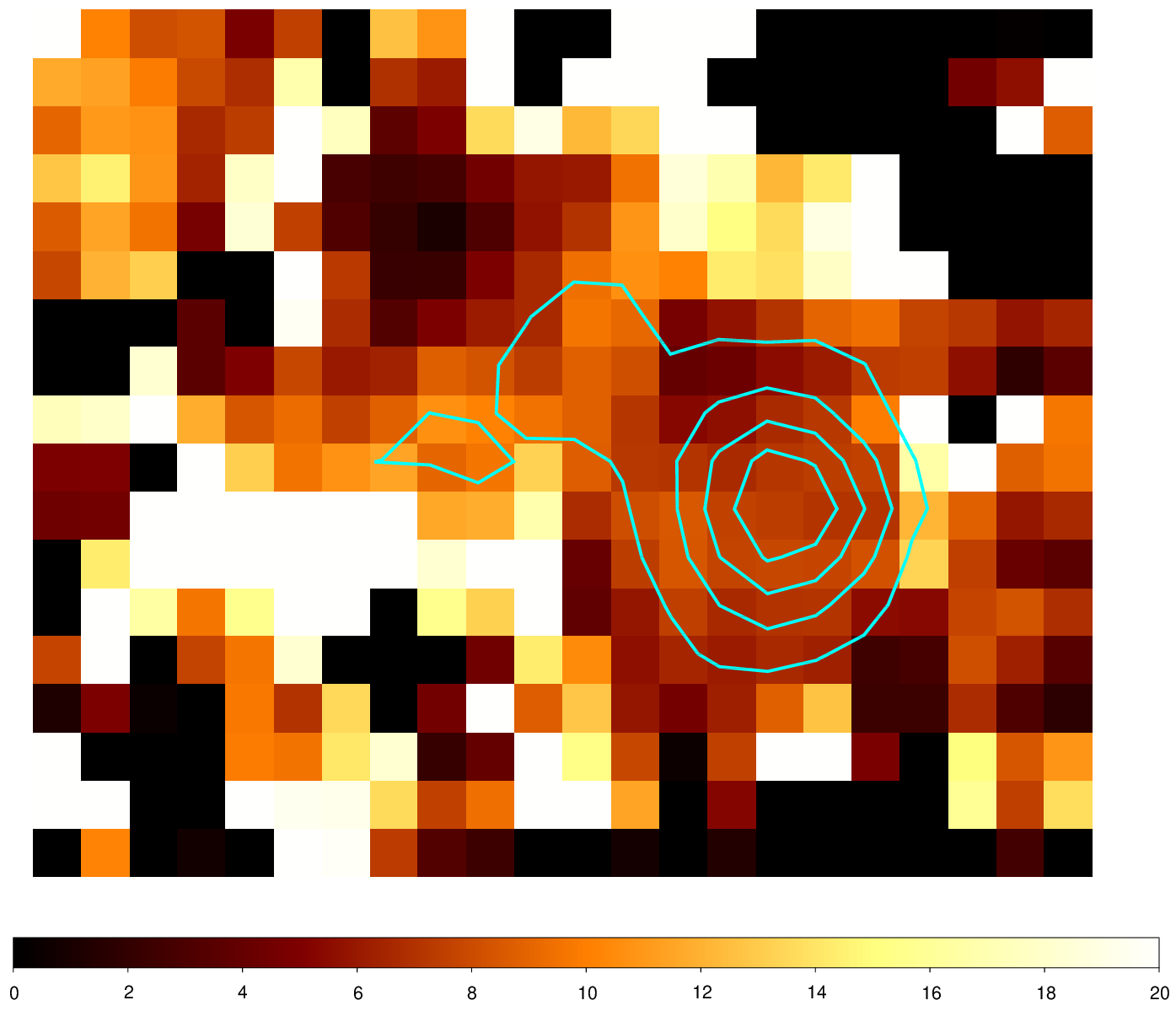}  
  \caption{$^{12}$CO/$^{13}$CO J=3--2 integrated line ratio for the N11B region, with contours of $^{13}$CO J=3--2.
}
  \label{N11B_div}
\end{figure}

\begin{figure}[!ht]
  \centering
  \includegraphics[width=0.7\textwidth]{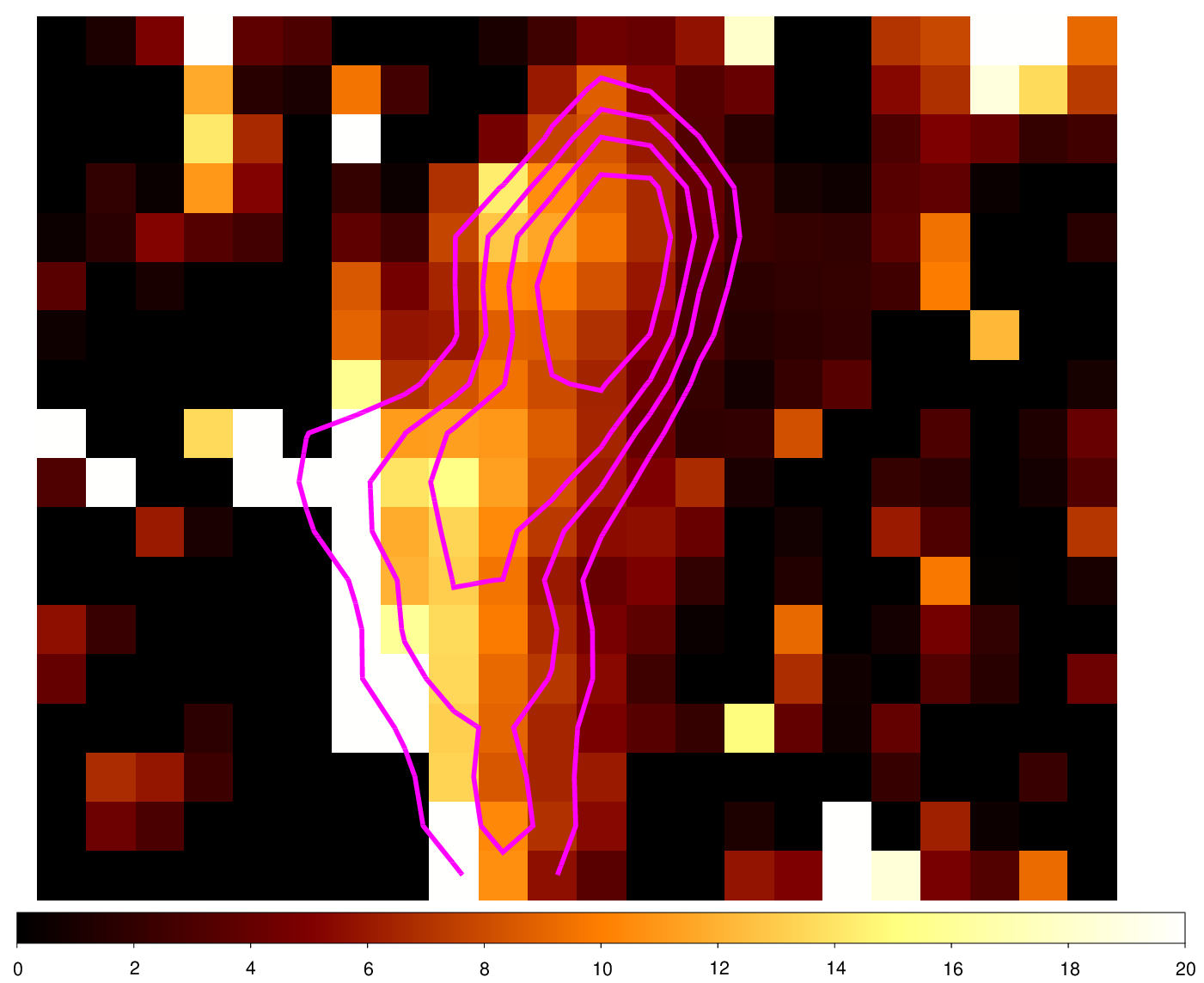}  
  \caption{$^{12}$CO/$^{13}$CO J=3--2 integrated line ratio for the N11B region, with contours of $^{12}$CO J=3--2.}
  \label{N11D_div}
\end{figure}

\begin{figure}[!ht]
  \centering
  \includegraphics[width=0.6\textwidth]{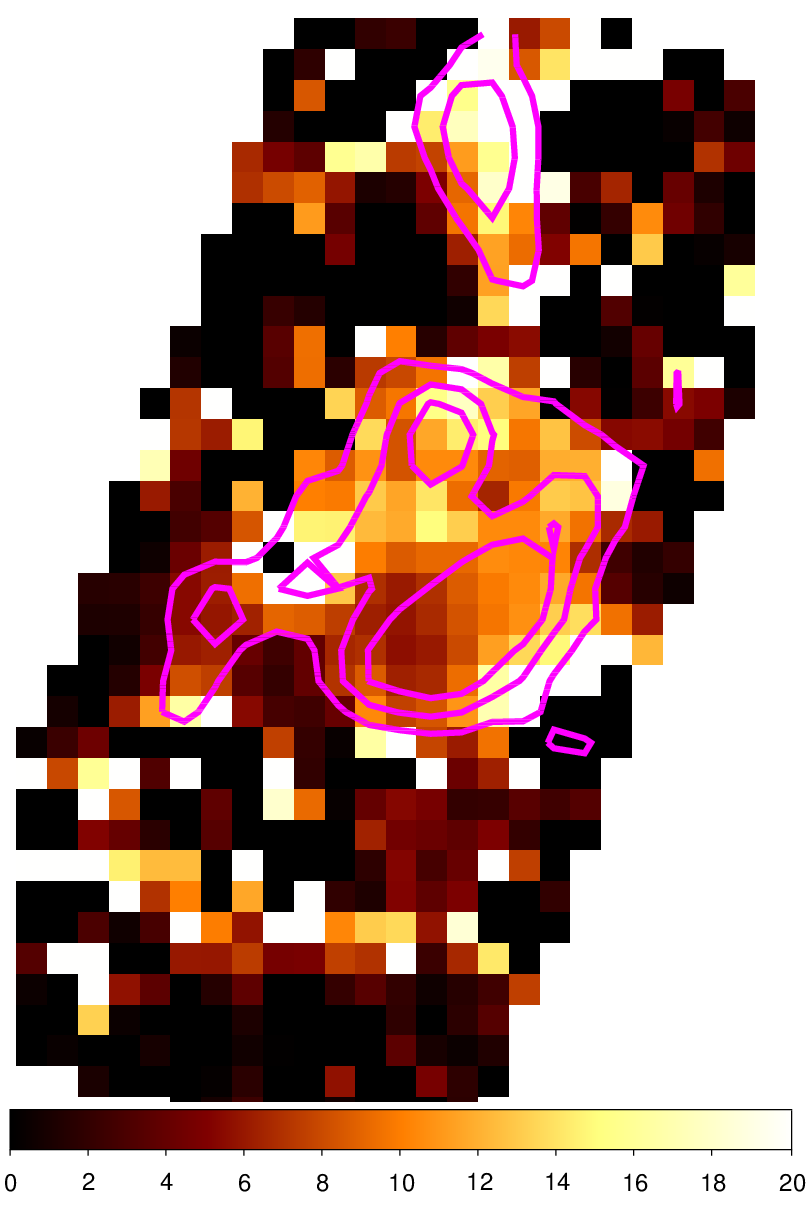}  
  \caption{$^{12}$CO/$^{13}$CO J=3--2 integrated line ratio for the N11I region, with contours of $^{12}$CO J=3--2.}
  \label{N11I_div}
\end{figure}

In the Figures. We also included maps of the integrated line ratio $^{12}$CO/$^{13}$CO for each region (see Figures \ref{N11B_div}, \ref{N11D_div} and \ref{N11I_div}). In the Figures it is also included contours of the integrated $^{12}$CO J=3--2 for reference. The average ratios obtained towards the regions delimited by the $^{12}$CO contours are 10 in N11B, and 8 in N11D and N11I.

These results are in agreement with values obtained towards other regions in the LMC \citep{Israel2003,Israel1991} and are somewhat larger than the typical values measuered in our Galaxy \citep{Burton2013}.

\subsection{Molecular mass of the studied regions}

Taking into account the morphology of each molecular cloud observed in the $^{13}$CO J=3-2 line (Figures \ref{mapasN11B}, \ref{mapasN11D}, and \ref{mapasN11I}), and assuming local thermodinamic equilibrium (LTE), we roughly estimate the mass of each cloud. Using the typical LTE formulae (see for example \citealt{Bu2010}) and assuming an isotopic abundance ratio X = 50 \citep{Wang2009} we estimate the $^{13}$CO column densities for each cloud. Then assuming $[H_2/^{13}$CO]=$1.8\times10^6$ \citep{Garay2002} we derived the corresponding $H_2$ column densities, from which, finally the cloud masses were estimated. The parameters obtained and used in this procedure are presented in Table \ref{table1}: the optical depths ($\tau_{12}$ and $\tau_{13}$), excitation temperatures (T$_{ex}$), $^{13}$CO column density (N($^{13}$CO)), and the cloud masses. The masses presented in Table \ref{table1} for the regions N11B, N11D and N11I are approximately 10$^4$ M$_\odot$, which are consistent with the results obtained by \citet{Herrera2013}.

\begin{table}[!t]
\centering
\caption{Physical parameters of the mapped clouds.}
\begin{tabular}{lccc}
\hline\hline\noalign{\smallskip}
\!\! & \!\!\!\!N11B & \!\!\!\!N11D & \!\!\!\!N11I \!\!\!\!\\
\hline\noalign{\smallskip}
\!\!$\tau_{12}$ 			  & 9.5    	      & 10.6 		   & 8.8\\
\!\!$\tau_{13}$ 			  & 0.1    		  & 0.2 		   & 0.1\\
\!\!T$_{ex}$ [K]			  & 12.9  		  & 11.2 		   & 9.7 \\
\!\!N($^{13}$CO) [cm$^{-2}$] & 1.3$\times10^{16}$ & 1.4$\times10^{16}$ & 2.2$\times10^{16}$\\
\!\!Mass(H$_2$) [M$_\odot$]  & 1.1$\times10^4$    & 1.3$\times10^4$    & 2.0$\times10^4$ \\

\hline
\end{tabular}
\label{table1}
\end{table}

\section{Acknowledgement}
This work was partially supported by Argentina grants awarded by UBA (UBACyT), CONICET and ANPCYT. 
M.R. wishes to acknowledge support from CONICYT(CHILE) through FONDECYT grant No1140839 and partial support from project BASAL PFB-06. M.C acknowledges support from FONDECYT grant No1140839. 
%


\bibliographystyle{plainnat}
\small
\bibliography{referencias}
 
\end{document}